\documentclass[%
    reprint,
    nobibnotes,
    amsmath,amssymb,
     aps,
    pre,
    floatfix,
   ]{revtex4-2}

\usepackage{graphicx}
\usepackage{bm}
\usepackage{xcolor}
\usepackage{color}
\usepackage{mathrsfs}

\usepackage{ragged2e}
\makeatletter
\long\def\@makecaption#1#2{%
  \vskip\abovecaptionskip
  \parbox{\columnwidth}{%
    \justifying
    \noindent #1.~#2\par
  }%
  \vskip\belowcaptionskip
}
\makeatother

\makeatletter
\def\check@deferlist@stuck#1{%
 \@ifx{\@deferlist@postshipout\@empty}{}{%
  \@ifx{\@deferlist@postshipout\@deferlist}{%
   \clearpage@sw{}{%
    \force@deferlist@stuck#1%
   }%
  }{}%
 }%
}%
\makeatother

\makeatletter
\providecommand \@ifxundefined [1]{%
 \@ifx{#1\undefined}%
}
\providecommand \@ifnum [1]{%
 \ifnum #1\expandafter \@firstoftwo
 \else \expandafter \@secondoftwo
 \fi
}
\providecommand \@ifx [1]{%
 \ifx #1\expandafter \@firstoftwo
 \else \expandafter \@secondoftwo
 \fi
}

\providecommand \href@noop [0]{\@secondoftwo}
\providecommand \href [0]{\begingroup \@sanitize@url \@href}
\providecommand \@href[1]{\@@startlink{#1}\@@href}
\providecommand \@@href[1]{\endgroup#1\@@endlink}
\providecommand \@sanitize@url [0]{\catcode `\\12\catcode `\$12\catcode
  `\&12\catcode `\#12\catcode `\^12\catcode `\_12\catcode `\%12\relax}
\providecommand \@@startlink[1]{}
\providecommand \@@endlink[0]{}
\providecommand \url [0]{\begingroup\@sanitize@url \@url}
\providecommand \@url [1]{\endgroup\@href {#1}{\urlprefix}}
\providecommand \urlprefix [0]{URL }

\providecommand \selectlanguage [0]{\@gobble}
\providecommand \bibinfo [0]{\@secondoftwo}
\providecommand \bibfield [0]{\@secondoftwo}

\providecommand \BibitemShut [1]{\csname bibitem#1\endcsname}
\let\auto@bib@innerbib\@empty
\makeatother

\begin{document}

\title{Velocity of an interface driven by an entropic force under shear flow}
\author{Yutaro Kado and Shin-ichi Sasa}
\affiliation{Department of Physics, Kyoto University, Kyoto 606-8502, Japan}
\date{\today}

\begin{abstract}
Thermal fluctuations can drive an interface when bulk fluctuation amplitudes differ between two phases. 
We study how shear flow modifies this mechanism using a stochastic non-conserved order-parameter model. 
For a planar interface parallel to the imposed shear, we derive its propagation velocity in the weak-noise and small-bias regime. 
The driving force comprises a shear-modified entropic contribution determined by bulk fluctuation spectra and a non-equilibrium contribution arising from the breaking of time-reversal symmetry.
At low shear rates, the entropic term provides the dominant contribution to the velocity of an interface near the zero-velocity plane of the shear flow,
and the shear-induced change in the velocity scales as the $4/3$ power of the shear rate.
At high shear rates, two-dimensional simulations show that the measured velocity is largely accounted for by the modified bulk entropic contribution.
\end{abstract}

\maketitle

\section{Introduction}

Non-equilibrium systems exhibit a wide range of interfacial phenomena, including domain-wall motion under heat flow \cite{Hinzke,Hals,Chico}, propagating reaction fronts \cite{Fisher,Kolmogorov,Saarloos}, shear-induced phase transitions \cite{Hess, Doi, Kuzuu,Olmsted_1990,Olmsted_1992,Olmsted_1997,Olmsted_1997,Pujolle-Robic,Mendil-Jakani,D_Parisi}, active matter \cite{Chate, Huber_2018}, and motility-induced phase separation \cite{Cates_2015,Cates_2025}.
In systems under shear flow, experiments on colloidal systems \cite{Holmqvist,Wu_2009,Shereda} and polymeric systems \cite{Sun} have investigated a variety of processes, including nucleation, interfacial growth, and melting or dissolution, with shear-induced acceleration observed in some cases.
Motivated by these experimental observations, an important question is whether interfacial propagation under shear admits a thermodynamic description and how the shear rate, as a non-equilibrium control parameter, enters the corresponding law of motion.

To approach these questions, we focus on the suppression of thermal fluctuations by shear flow as a possible mechanism affecting interfacial motion.
For example, in sheared systems, anisotropic deformation of fluctuation correlations \cite{De_Gennes_1976, Safinya}, shifts in critical temperatures \cite{Onuki,Beysens_1980,Beysens_1983},
shear-induced ordering and hydrodynamic phenomena in $p$-atic liquid crystals \cite{Giomi_Lett,Giomi_E,Krommydas},
the emergence of long-range order in two-dimensional systems \cite{Nakano}, changes in the upper critical dimension \cite{Ikeda}, and shear-induced pressure anomalies \cite{Kawasaki,Wada} have been reported.
Analyses based on order-parameter dynamics have established that fluctuations in a single phase depend systematically on the shear rate \cite{Onuki,Nakano}.

We therefore ask how the shear-induced modification of bulk fluctuations affects the propagation of an interface between two coexisting phases.
A key ingredient in our analysis is the mechanism of fluctuation-induced interface propagation reported in Refs. \cite{Costantini, Kado}.
In a stochastic non-conserved order-parameter model for equilibrium dynamics, bulk thermal fluctuations contribute an entropic term to the free energy.
Because this contribution is proportional to temperature and is determined by the difference in fluctuations between the two bulk phases, it acts as an effective driving force on an interface.
We refer to this noise-induced driving force as the entropic force.
Here and below, “entropic” refers to such a fluctuation contribution to the effective potential.

In this study, we investigate how the shear-induced suppression of bulk fluctuations modifies the propagation of the interface driven by the entropic force.
The interface-velocity formula contains two distinct driving contributions: a shear-modified entropic force and an additional force arising from the breaking of time-reversal symmetry in the non-equilibrium system.
For small shear rates, the entropic contribution dominates, and the velocity of the interface near the zero-shear-velocity plane has a correction proportional to the $4/3$ power of the shear rate.
Numerical calculations in two dimensions show that, at large shear rates, the measured shear-rate dependence of the velocity is largely captured by the shear-modified bulk entropic contribution.
The analytical and numerical results suggest that shear flow alters the propagation velocity through the entropic force.

The remainder of the paper is organized as follows.
Section II introduces the model and reviews the basic mechanism of interface motion driven by the entropic force.
Section III presents the main results for the interfacial velocity under shear flow.
Section IV presents the derivation.
Section V summarizes the results and discusses future directions.

\section{Preliminaries}

\subsection{Model}

Let ${\bm r}=(x_1, \cdots,x_d)\in \mathbb{R}^d$ denote a position in the $d$-dimensional region $D\equiv[-L_1, L_1]\times[0,L]^{d-1}$, where $d \geq 2$.
Both $L_1$ and $L$ are sufficiently large, and $L$ is taken to be infinite in the theoretical analysis.
We also introduce the notation ${\bm r}_{\perp}=(x_2, \cdots,x_d)$.
We define a real scalar order parameter field $\phi({\bm r}, t)$ in the region $D$.
The free-energy functional of $\phi$ is given by
\begin{eqnarray}
{\cal F}(\phi)\equiv \int_{D} d{\boldsymbol r} \Big[f(\phi)
+ \frac{\kappa}{2}\sum_{i=1}^d (\partial_{x_i}\phi)^2 \Big],
\end{eqnarray}
where $f(\phi)$ is the mesoscopic free energy density and $\kappa$ is a constant characterizing the interfacial energy.
The order parameter is convected by a steady uniform shear flow with velocity ${\bm v}({\bm r}) = (0, \dot{\gamma}x_1,0,\cdots,0)$, where $\dot{\gamma}\geq0$ without loss of generality.
Following Onsager's principle, we assume that the dynamics of $\phi({\bf r},t)$ is described by
\begin{eqnarray}
\Big(\partial_{t} + {\bm v}({\bf r}) \cdot \nabla \Big)\phi({\bm r},t) = -\Gamma \frac{\delta {\cal F}}{\delta \phi}
  + \eta,
\label{dynamics}
\end{eqnarray}
where $\Gamma$ is a constant representing the mobility and $\eta$ is Gaussian white noise satisfying $\langle \eta({\boldsymbol r},t)\rangle=0$ and the fluctuation-dissipation relation
\begin{eqnarray}
\langle \eta({\boldsymbol r},t) \eta({\boldsymbol r'},t') \rangle = 2\Gamma T\delta({\boldsymbol r}-{\boldsymbol r'})\delta(t-t').
\end{eqnarray}
This model is an extension of “Model A” \cite{Hohenberg_Halperin}, which describes the dynamics of a non-conserved order parameter.
Because this model describes mesoscopic dynamics, we introduce a microscopic cutoff region $D_k$ in wave-number space; Fourier modes of $\phi$ with $k \notin D_k$ are set to zero.
We consider a system in which the mesoscopic free-energy density $f(\phi)$ has two minima at $\phi_1$ and $\phi_2$. Without loss of generality, we assume $\phi_1<\phi_2$.

We impose periodic boundary conditions in the ${\bm r}_\perp$ directions and fixed boundary conditions $\phi(-L_1, {\bm r}_\perp) = \phi_1$ and $\phi(L_1, {\bm r}_\perp) = \phi_2$.
As the initial condition, a planar interface is placed at $x_1=0$.
We restrict our analysis to timescales shorter than those associated with nucleation in the bulk regions.

\subsection{Propagation velocity of a planar interface}

We review the expression for the propagation velocity of a planar interface in the absence of shear, $\dot{\gamma}=0$. We first consider the deterministic case $T=0$ and then discuss the case $T>0$, in which a fluctuation-induced driving force appears.

\subsubsection{Deterministic case: $T=0$}
When $T=0$, the interface moves in the $x_1$ direction owing to the difference in free-energy density $f(\phi)$ between the two bulk regions.
A steadily propagating solution $\phi_0(w)$ with $w\equiv x_1-c_{\mathrm{det}}t$ satisfies
\begin{eqnarray}
-c_{\mathrm{det}}\partial_w \phi_0(w)=-\Gamma[f'(\phi_0)-\kappa\partial_w^2\phi_0],
\label{T0}
\end{eqnarray}
where $c_{\mathrm{det}}$ is the steady propagation velocity of the interface.
Mathematically, Eq. (\ref{T0}) is a nonlinear eigenvalue problem in which the profile $\phi_0$ and the velocity $c_{\mathrm{det}}$ are determined simultaneously.
Taking the limit $L_1\to \infty$, we derive a formula relating the propagation velocity $c_{\mathrm{det}}$ to the profile $\phi_0$.
Multiplying Eq. (\ref{T0}) by $\partial_w\phi_0$ and integrating over $[-\infty,\infty]$ with respect to $w$, we obtain
\begin{eqnarray}
c_{\mathrm{det}}=\frac{\Gamma \delta f}{\int_{-\infty}^{\infty} dw ( \partial_{w}\phi_0(w) )^2}.
\end{eqnarray}
Here, $\delta f \equiv f(\phi_2) - f(\phi_1)$,
which is the free-energy difference across the interface and therefore represents the interfacial driving force.
The prefactor multiplying $\delta f$ is the interfacial mobility.

\subsubsection{Interface driven by an entropic force: $T>0$}

We now consider the case in which the free-energy difference $\delta f$ between the two phases is small.
We write $f(\phi)$ as
\begin{eqnarray}
f(\phi)=f_0(\phi)-h\phi.
\label{f0}
\end{eqnarray}
Here, $f_0$ has degenerate minima at $\phi_1$ and $\phi_2$, i.e., $f_0(\phi_1)=f_0(\phi_2)$, and $h$ is sufficiently small.

\begin{figure}[tbp]
  \centering

    \includegraphics[width=\linewidth]{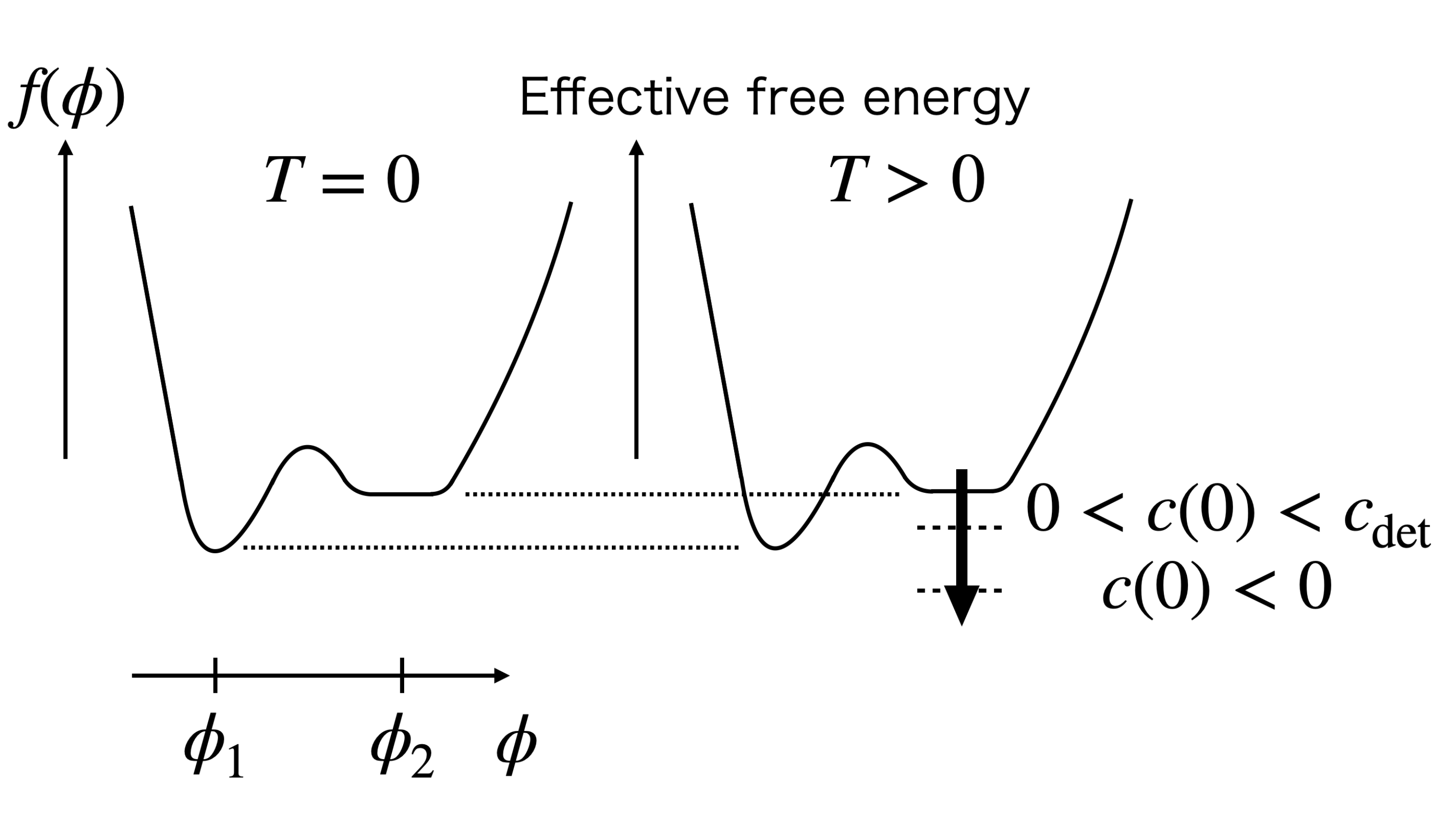}

\caption{
Schematic of the potential shifted by thermal fluctuations when $\xi_1<\xi_2$.
The left panel represents the free energy $f(\phi)$.
The right panel represents the effective free energy including the fluctuation contribution, whose minima have values $f(\phi_i)-Ts_i$, $i=1,2$.
}
\label{fig:image1}
\end{figure}

We treat thermal noise and $\delta f$ as perturbations around $\varphi_0$, which solves
\begin{eqnarray}
f_0'(\varphi_0)-\kappa\partial_w^2\varphi_0=0.
\end{eqnarray}
Let $\theta(\bm{r}_{\perp}, t)$ denote the $x_1$ coordinate of the interface at time $t$.
We compute the average propagation velocity $c(0)=c(\dot{\gamma}=0)=\langle \partial_t \theta(\bm{r}_{\perp}, t) \rangle|_{\dot{\gamma}=0}$.
The result is \cite{Costantini,Kado}
\begin{eqnarray}
c(0)= \Gamma_{\mathrm{int}} [\delta f -T\delta s] + O(T^{\frac{3}{2}},T^{\frac{1}{2} }h,h^2)
\label{c0}
\end{eqnarray}
with
\begin{eqnarray}
\Gamma_{\mathrm{int}}&\equiv&\frac{\Gamma}{ \int_{-\infty}^{\infty} dw ( \partial_{w}\varphi_0(w) )^2}.
\end{eqnarray}
Here, $\Gamma_{\mathrm{int}}$ is the mobility of the planar interface.
$\delta f = f(\phi_2) - f(\phi_1) \simeq - h \Delta \phi$ is the free-energy difference between the two phases, and $\Delta \phi \equiv \phi_2-\phi_1$.
$T\delta s \equiv T(s_2 - s_1)$ is the effective free energy difference associated with bulk fluctuations, where
\begin{eqnarray}
s_i&=&-\frac{1}{2}\int_{D_k} \frac{d{\bm k}}{(2\pi)^d} \frac{\xi_i^{-2}-k_1^2+\displaystyle \sum_{j=2}^{d}k_j^2}{\xi_i^{-2}+\bm{k}^2}, \quad i=1,2.
\end{eqnarray}
Here, $\xi_i=\sqrt{\kappa/f_0^{''}(\phi_i)}$, $i=1,2$, are the correlation lengths of the fluctuations in the bulk regions.
We refer to this noise-induced contribution and the associated driving force as the entropic contribution and the entropic force.
Thus, a nonzero $\delta s$ requires $\xi_1\neq\xi_2$.
A larger curvature at a minimum corresponds to stronger local confinement and changes the entropic contribution of the corresponding bulk phase.
For example, $\xi_1<\xi_2$ implies $s_1<s_2$.
The change in velocity can therefore be interpreted as a fluctuation-induced shift of the effective free energy, as illustrated in Fig. \ref{fig:image1}.
When $\xi_1<\xi_2$ and $\delta f>0$,
the velocity $c(0)$ can be smaller than $c_{\mathrm{det}}$, and the direction of motion can be reversed in some parameter regimes.
Conversely, if $\delta f<0$, the interface speed $|c(0)|$ becomes larger than $|c_{\mathrm{det}}|$.
Depending on the competition between $\delta f$ and $T\delta s$, the direction of motion can be reversed.

For $d=1$, taking $D_k=[-k_c,k_c]$ and $k_c \to \infty$ gives
\begin{eqnarray}
s_2-s_1=-\frac{1}{2}\Big(\frac{1}{\xi_2}-\frac{1}{\xi_1} \Big).
\end{eqnarray}
This formula for $d=1$ was first derived in Ref. \cite{Costantini}.
For $d=2$, if we take $D_k$ to be a spherical region, i.e., $D_k=\{{\bm k}\mid |{\bm k}|<k_{c,\mathrm{sph}} \}$, we obtain \cite{Kado}
\begin{eqnarray}
s_i=-\frac{1}{8 \pi \xi_i^2}{\rm ln}(\xi_i^2 k_{c,\mathrm{sph}}^2+1).
\label{s_2dim}
\end{eqnarray}
In general, for $d \geq 2$, $c(0)$ depends on the cutoff wavenumber $k_{c,\mathrm{sph}}$.

\section{Main results}

We now consider how the entropic-force mechanism is modified under shear flow.
Section III.A presents the general velocity formula under shear flow and identifies its two driving contributions.
Section III.B analyzes the low-shear-rate regime and derives the leading shear-rate dependence of the propagation velocity of the interface near $x_1=0$.
Section III.C reports numerical calculations in $d=2$ for the high-shear-rate regime
and quantifies the deviation of the measured propagation velocity from the theoretical prediction based on the shear-modified bulk entropic contribution.

\subsection{Velocity under shear flow}

Using the potential in Eq. (\ref{f0}) and assuming that $T$ and $h$ are small, we obtain the shear-dependent velocity $c(\dot{\gamma}) \equiv \langle \partial_t \theta(\bm{r}_{\perp}, t) \rangle$:
\begin{equation}
c(\dot{\gamma})
=
\Gamma_{\mathrm{int}}
[\delta f-T\delta s(\dot{\gamma})+g_{\mathrm{neq}}]
+O(T^{3/2},T^{1/2}h,h^2).
\label{result}
\end{equation}
The derivation of Eq. (\ref{result}) is presented in Sec. IV.
The term $\delta f -T \delta s(\dot{\gamma})$ in Eq. (\ref{result}) represents the driving force associated with the shear-modified Gaussian fluctuation contribution from the bulk phases.
Here, $T \delta s(\dot{\gamma})=T[s_2(\dot{\gamma})-s_1(\dot{\gamma})]$, and
\small
\begin{eqnarray}
-T s_i(\dot{\gamma})&\equiv& \frac{T}{2} \int_{D_k}\frac{d{\bm k}}{(2\pi)^d} \notag\\
&&\times
\frac{\xi_i^{-2}-k_{1}^2+\displaystyle \sum_{j=2}^{d}k_{j}^2}{\xi_i^{-2}+{\bm k}^2+a_0(\Gamma^{-1}\kappa^{-1}|\dot{\gamma}||k_{2}|)^{\frac{2}{3}}}.
\label{s_form}
\end{eqnarray}
\normalsize
Here, $a_0\equiv[12^{1/3}\Gamma(4/3)]^{-1}\simeq0.49$ and $D_k=[-k_{c,\mathrm{cub}},k_{c,\mathrm{cub}}]^d$.
The shear-dependent entropic contribution originates from order-parameter fluctuations in the bulk regions.
As in Ref. \cite{Kado}, $s_i(\dot{\gamma})$ depends on the cutoff $k_{c,\mathrm{cub}}$.
The integrand in Eq. (\ref{s_form}) was given in Ref. \cite{Nakano} as an approximate expression for the fluctuation correlation function in a single-phase system under shear flow.
The second term, $g_{\mathrm{neq}}$, is given by
\begin{eqnarray}
g_{\mathrm{neq}} &\equiv& \frac{1}{\Gamma L^{d-1}} \int_{[0,L]^{d-1}} d{\bm r}_{\perp} \int_{-\infty}^{\infty} dw \notag \\
&&\times \left\langle \partial_w \rho_1
\left(\partial_{t}+\dot{\gamma}(w+\theta({\bm r}_{\perp},t))\partial_{x_2}\right)\rho_1\right\rangle.
\end{eqnarray}
Here, $\rho_1(w, \bm{r}_{\perp} )\equiv\phi(\bm{r})-\varphi_0(w)$ is the order-parameter fluctuation, 
and $w$ is a comoving coordinate defined by $w=x_1- \theta(\bm{r}_{\perp},t)$.
In contrast to $\delta s(\dot{\gamma})$, which is determined by homogeneous bulk fluctuation spectra,
$g_{\mathrm{neq}}$ collects the residual contributions from fluctuations in the interfacial region.
Furthermore, $g_{\mathrm{neq}}$ is a genuinely non-equilibrium contribution associated with time-reversal symmetry breaking: it vanishes in the comoving equilibrium system at $\dot{\gamma}=0$ because of time-reversal symmetry \cite{Kado}.

\subsection{Low-shear-rate regime}

We next analyze how $\delta s(\dot{\gamma})$ and $g_{\mathrm{neq}}$ depend on the shear rate. Together, these determine the leading shear-rate dependence of $c(\dot{\gamma})$ in the weak-shear limit.

\subsubsection{Entropic force}
\label{d_s_gamma}

From the denominator of Eq. (\ref{s_form}), the low-shear condition is
\begin{eqnarray}
a_0(\Gamma^{-1}\kappa^{-1}|\dot{\gamma}||k_{2}|)^{2/3}
&\ll& \xi_i^{-2}+k_2^2, \quad i=1,2,
\label{g1_def}
\end{eqnarray}
for all values of $k_2$.
Requiring the two sides of Eq. (\ref{g1_def}), viewed as functions of $k_2$, to have no intersection gives
\begin{eqnarray}
\dot{\gamma} \ll \dot{\gamma}_1 \equiv
\frac{1}{2} \Big(\frac{3}{a_0}\Big)^{\frac{3}{2}} \frac{\Gamma \kappa}{{\rm max}\{\xi_1,\xi_2\}^2}.
\end{eqnarray}
Under this condition, $s_i(\dot{\gamma})$ can be expanded as
\begin{eqnarray}
-s_i(\dot{\gamma}) = -s_i(0)-A_{i,1}|\dot{\gamma}|^{\frac{2}{3}}
+A_{i,2}|\dot{\gamma}|^{\frac{4}{3}} +O(|\dot{\gamma}|^2).
\end{eqnarray}
Here, $s_i(0)$, $A_{i,1}$, and $A_{i,2}$ are independent of $\dot{\gamma}$ and are given by wave-number integrals.
Specifically,
\begin{eqnarray}
A_{i,n} \equiv \frac{
\left[a_0(\Gamma\kappa)^{-\frac{2}{3}}\right]^n
k_{c,\mathrm{cub}}^{d-\frac{4}{3}n} }{2(2\pi)^d} \times I_n(\xi_i k_{c,\mathrm{cub}})
\end{eqnarray}
with
\begin{eqnarray}
I_n(x) &\equiv& x^{-d+\frac{4}{3}n}\int_{[-x,x]^d} d{\bm u}
\frac{|u_2|^{\frac{2}{3}n}}{(1+{\bm u}^2)^{1+n}} \notag \\
&&\times \left(1-u_{1}^2+\sum_{j=2}^{d}u_{j}^2\right).
\end{eqnarray}
In the limit $x \to \infty$, $I_1(x)$ converges to a finite value.
Because the model is mesoscopic, we assume that $\xi_{i} k_{c,\mathrm{cub}} \gg 1$.
Thus, taking $\xi_i k_{c,\mathrm{cub}} \to \infty$, we find $A_{1,1}=A_{2,1}$.
The leading $|\dot{\gamma}|^{2/3}$ correction is therefore independent of the bulk correlation length in this mesoscopic limit and contributes equally to the two phases.
As a result, this term cancels in the difference $\delta s(\dot{\gamma}) = s_2(\dot{\gamma}) - s_1(\dot{\gamma})$,
and the first phase-dependent correction is the $|\dot{\gamma}|^{4/3}$ term.
Since $I_2(x)$ is an increasing function of $x$, the sign of
$A_{2,2}-A_{1,2}$ is determined by the sign of $\xi_2-\xi_1$.
Thus, in general, we can write $A_{2,2}-A_{1,2}=\operatorname{sgn}(\xi_2-\xi_1)\,|A_{2,2}-A_{1,2}|$.
Therefore, we obtain
\begin{eqnarray}
-\delta s(\dot{\gamma})&=&-\delta s(0) \notag \\
&+& \operatorname{sgn}(\xi_2-\xi_1)\,|A_{2,2}-A_{1,2}| |\dot{\gamma}|^{\frac{4}{3}}+\cdots.
\end{eqnarray}
Thus, $|\delta s(\dot{\gamma})|<|\delta s(0)|$, which shows that the magnitude of the entropic force decreases under shear flow.

\subsubsection{Contribution due to the breaking of time-reversal symmetry}

\begin{figure}[tbp]
    \centering
    \includegraphics[width=\columnwidth]{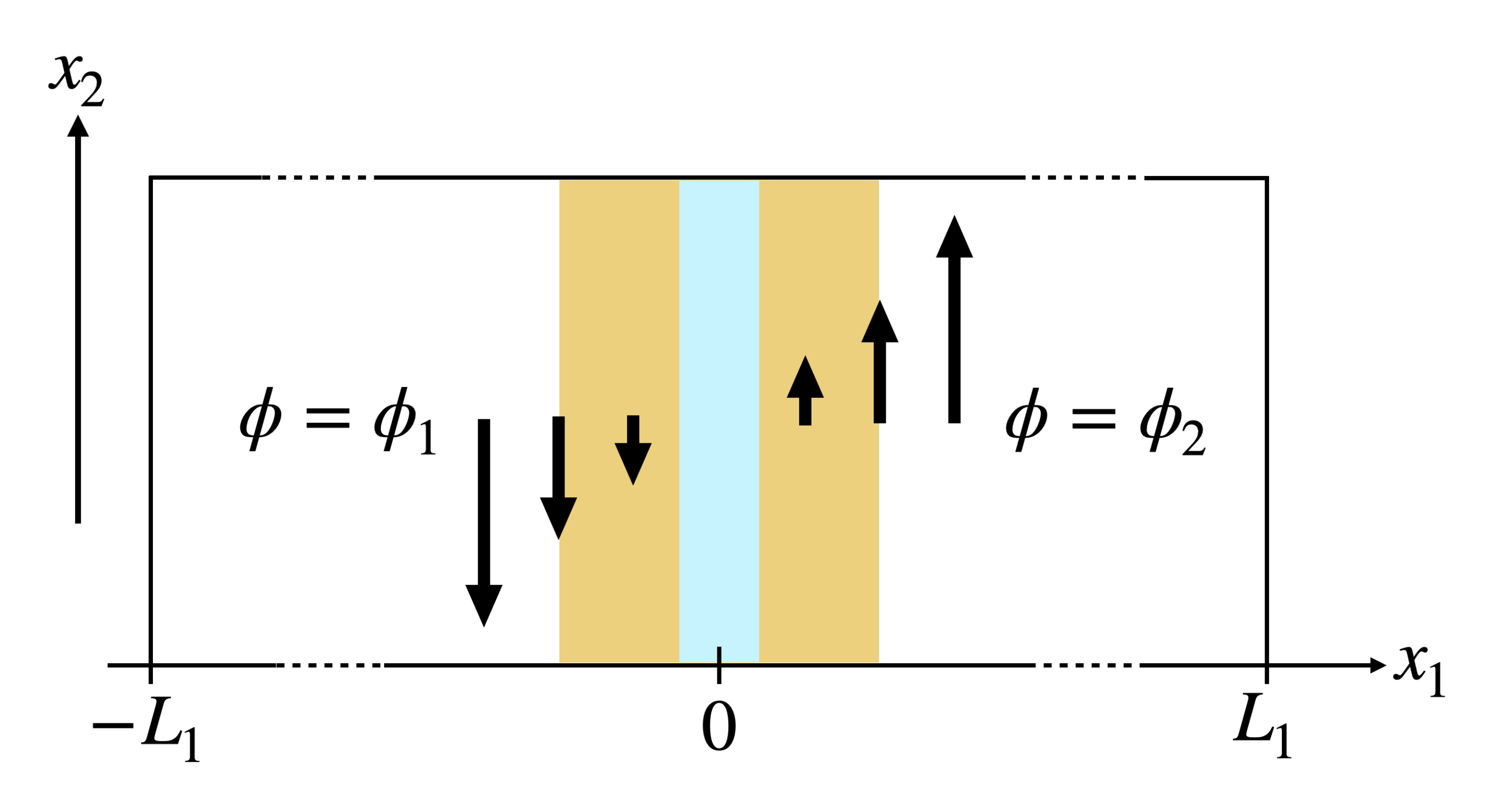}
\caption{
Schematic in the $x_1$-$x_2$ plane.
The two sides of the interface correspond to the $\phi=\phi_1$ and $\phi=\phi_2$ phases, respectively, separated by the interfacial region, which is represented by the blue region.
The orange region is centered at $x_1=0$, and its length is on the order of several times the interfacial width.
}
\label{fig:image2}
\end{figure}

When the interface lies in the interfacial-scale region centered at $x_1=0$, schematically shown in orange in Fig. \ref{fig:image2}, and the shear rate is small, we examine the shear-rate dependence of $g_{\mathrm{neq}}$.
From Eq. (\ref{rho_1_time}) in Appendix \ref{app:derivation}, the integrand of $g_{\mathrm{neq}}$ is nonzero only near $w=0$, which corresponds to the vicinity of the interface.
Thus, $g_{\mathrm{neq}}$ is determined by fluctuations in the interfacial region.
The relevant range of integration with respect to $w$ is therefore restricted to $[-K l_{\mathrm{int}},K l_{\mathrm{int}}]$, where $K$ is a positive constant of order unity.
We also consider the case in which the advection term of Eq. (\ref{dynamics}), which also appears in the dynamics of $\rho_1$ given by Eq. (\ref{rho1}), is smaller than the thermodynamic driving term in the equation.
This condition holds when $\dot{\gamma}$ satisfies
\begin{eqnarray}
\frac{\dot{\gamma} K l_{\mathrm{int}}}{\xi_i} \ll \Gamma f^{''}(\phi_i), \quad i=1,2,
\end{eqnarray}
i.e.,
\begin{eqnarray}
\dot{\gamma} \ll \dot{\gamma}_2 \equiv \frac{\Gamma \kappa}{K l_{\mathrm{int}} {\rm max}\{\xi_1,\xi_2\}},
\end{eqnarray}
Here, $l_{\mathrm{int}}$ is the interfacial width.
When these conditions are satisfied, $\rho_1$ can be expanded in $[-K l_{\mathrm{int}},K l_{\mathrm{int}}]\times[-L,L]^{d-1}$ as
\begin{eqnarray}
\rho_1(w,{\bm r}_{\perp} ,t)&=&\rho_{1,0}(w,{\bm r}_{\perp},t) \notag \\
&+&\Big(\frac{\dot{\gamma} }{\dot{\gamma}_2}\Big)\rho_{1,1}(w,{\bm r}_{\perp},t)+\cdots.
\end{eqnarray}
Substituting this expression into $g_{\mathrm{neq}}$, we can expand $g_{\mathrm{neq}}$ in powers of $\dot{\gamma}$.
At $\dot{\gamma}=0$, $g_{\mathrm{neq}}$ vanishes by time-reversal symmetry of the equilibrium system \cite{Kado}.
The term linear in $\dot{\gamma}$ is also forbidden by the symmetry under the simultaneous transformation
$x_2\to -x_2$ and $\dot{\gamma}\to -\dot{\gamma}$. 
Hence, at small shear rates,
\begin{eqnarray}
g_{\mathrm{neq}} \propto \dot{\gamma}^2.
\label{g_neq_scaling}
\end{eqnarray}
We note that because the bulk correlation lengths and the interfacial width are of the same order, $\dot{\gamma}_2$ is also of the same order of magnitude as $\dot{\gamma}_1$.

\subsubsection{Velocity in the low-shear-rate regime}

When $ \dot{\gamma} \ll {\rm min}\{ \dot{\gamma}_1,\dot{\gamma}_2 \} $ and the interface is located near $x_1=0$, the interface velocity satisfies
\begin{eqnarray}
c(\dot{\gamma}) - c(0) &=& {\rm sgn}( \xi_2-\xi_1 )
\Gamma_{\mathrm{int}} T |A_{2,2}-A_{1,2}| \notag \\
&&\times |\dot{\gamma}|^{ \frac{4}{3} } + O(|\dot{\gamma}|^{2} ).
\label{low_shear}
\end{eqnarray}

\begin{figure}[tbp]
  \centering
    \includegraphics[width=\linewidth]{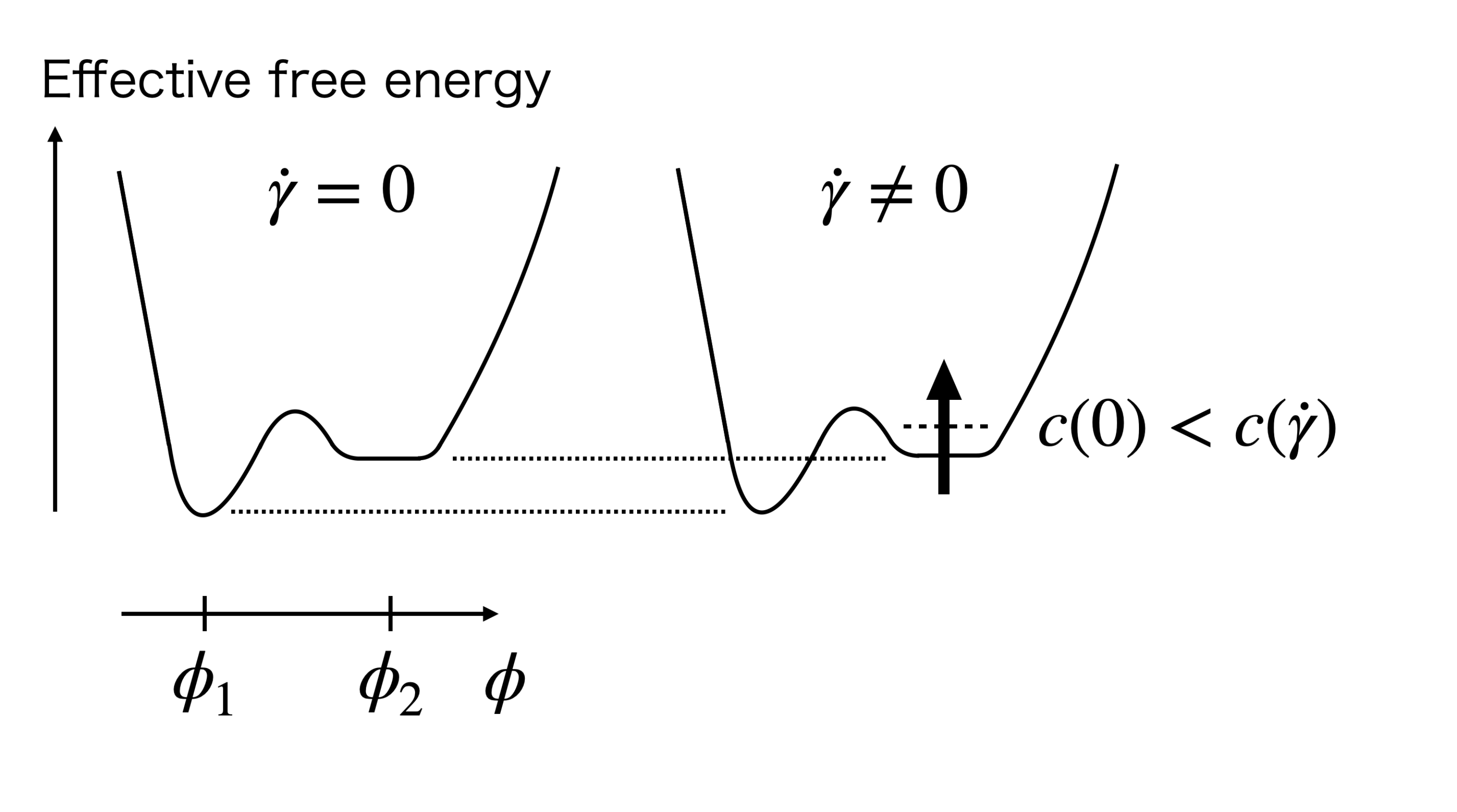}
\caption{
Schematic illustration of the shear-modified potential at a low shear rate, with the interface located near $x_1=0$ and $\xi_1<\xi_2$.
The left panel represents the effective free energy at $\dot{\gamma}=0$.
The right panel represents the shear-modified free energy, whose minima have values $f(\phi_i)-Ts_i(\dot{\gamma})$, $i=1,2$.
}
\label{fig:image3}
\end{figure}

Within the approximation of Eq. (\ref{s_form}), Eq. (\ref{low_shear}) gives the first phase-dependent shear correction proportional to $|\dot{\gamma}|^{4/3}$.
Therefore, if the shear rate is small and the interface is located near $x_1=0$, the potential picture can be extended, as illustrated in Fig. \ref{fig:image3}.
When $\xi_1<\xi_2$, the relation $\delta s(\dot{\gamma})< \delta s(0)$ holds, and thus $c(\dot{\gamma})>c(0)$.
In general, shear flow reduces the magnitude of the entropic force induced by thermal fluctuations, causing $c(\dot{\gamma})$ to approach the deterministic velocity $c_{\mathrm{det}}$.
This shift in the entropic contribution indicates that shear flow can change both the speed and the direction of propagation.

\subsection{Numerical results for the high-shear-rate regime}

Section III.B derives the propagation velocity of the interface near $x_1=0$ in the low-shear-rate regime, where $g_{\mathrm{neq}}$ is small.
By contrast, in the high-shear-rate regime, $ \dot{\gamma} \gg {\rm min}\{ \dot{\gamma}_1,\dot{\gamma}_2 \} $, $g_{\mathrm{neq}}$ cannot be evaluated analytically.
In the simulations below, we do not evaluate $g_{\mathrm{neq}}$ directly. Instead, we
compare the numerically measured propagation velocity with the bulk thermodynamic
contribution multiplied by the interfacial mobility,
\[
c_{\mathrm{thermo}}(\dot{\gamma})\equiv \Gamma_{\mathrm{int}}[\delta f-T\delta s(\dot{\gamma})].
\]
The purpose of this comparison is to determine whether deviations from
$c_{\mathrm{thermo}}(\dot{\gamma})$, which would include the contribution from $g_{\mathrm{neq}}$,
can be resolved within the present numerical accuracy.
We perform the numerical calculations for $d=2$.

\subsubsection{Numerical method}

We discretize the model on a square lattice with spatial mesh size $\Delta x$, chosen to be smaller than both $\xi_1$ and $\xi_2$.
The time step is denoted by $\Delta t$.
Appendix \ref{app:numerics} summarizes the computational methods for the time evolution.

The numerical calculations use the potential
\begin{eqnarray}
f_0(\phi) = A\Bigg( \frac{ 1-{\rm e}^{b_1(\phi-1)} }{ 1-{\rm e}^{b_1(\phi_0-1)} } \frac{ 1-{\rm e}^{-b_2(\phi+1)} }{ 1-{\rm e}^{-b_2(\phi_0+1)} }\Bigg)^2.
\label{f_0}
\end{eqnarray}
For this potential, $\phi_1=-1$ and $\phi_2=1$. By choosing the parameters appropriately, one can impose
$f_0(\phi_1)=f_0(\phi_2)$ while maintaining $f''_0(\phi_1)\neq f''_0(\phi_2)$.
For the initial condition in the numerical calculations, we use
\begin{eqnarray}
\phi({\bm r},t=0)= \frac{\phi_2-\phi_1}{2} {\rm tanh}\Big(\frac{x_1}{\sqrt{\kappa}}\Big) + \frac{\phi_1+\phi_2}{2}.
\end{eqnarray}

\subsubsection{Numerical results}

\begin{figure}[tbp]
  \centering
    \IfFileExists{t_v-eps-converted-to.pdf}{%
      \includegraphics[width=0.9\linewidth]{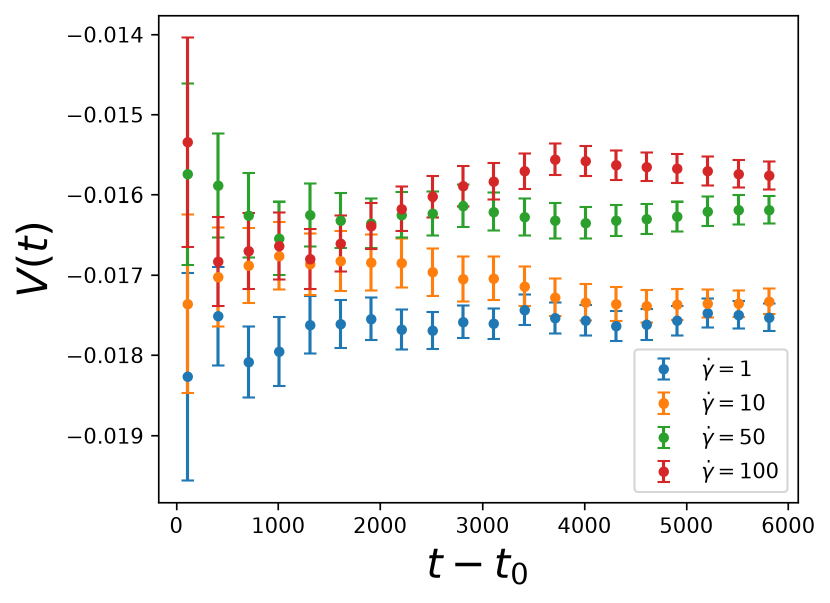}%
    }{%
      \fbox{\parbox[c][0.28\textheight][c]{0.86\linewidth}{\centering
      Placeholder for the missing figure file \texttt{t\_v.eps}.}}%
    }
    \caption{
    Ensemble average of $V(t)$ for several values of $\dot{\gamma}$.
    The grid spacing is $\Delta x=0.8$, the time step is $\Delta t=0.0005$, and the system sizes are $L_1=240$ and $L=160$.
    The physical parameters are $\kappa=1600$, $\Gamma=0.1$, $T=0.5$, $\xi_1=4.33$, $\xi_2=27.3$, and $h=0$.
    The parameters in Eq. (\ref{f_0}) are $A=1$, $b_1=0.5$, $b_2=5.0$, and $\phi_0=-0.5$.
    The velocity is calculated using 80 samples, with $t_0=50$.
    }
    \label{fig:image4}
\end{figure}

We calculate the interface profile averaged over the $x_2$ direction,
\begin{eqnarray}
\Phi(x_1,t) \equiv \frac{1}{L}\int_0^{L} d x_2 \phi(x_1,x_2,t)
\end{eqnarray}
and define the interface position $X(t)$ by $\Phi(X(t),t)=0$. The time-averaged velocity is then
\begin{eqnarray}
V(t)=\frac{X(t)-X(t_0)}{t-t_0}.
\end{eqnarray}
Figure \ref{fig:image4} shows the ensemble average of $V(t)$ over multiple simulation runs.
When the interface remains near $x_1=0$ under shear flow, its average velocity is constant over the observation time interval.
The plateau value gives the numerical estimate of $c(\dot{\gamma})$.

Because $\delta s(\dot{\gamma})$ and $c_{\mathrm{thermo}}(\dot{\gamma})$ depend on the cutoff, we must determine $k_{c,\mathrm{cub}}$.
A previous study \cite{Kado}  identified the cutoff wavenumber with the inverse of the discrete lattice spacing and fitted the theoretical expression in Eq. (\ref{s_2dim}) to the numerical values of $c(0)$,
using $k_{c,\mathrm{sph}}=2\sqrt{2}\pi/\Delta x$.
Therefore, we choose $k_{c,\mathrm{cub}}$ such that $c_{\mathrm{thermo}}(0)$ matches the theoretical value of $c(0)$ for this choice of $k_{c,\mathrm{sph}}$.

\begin{figure}[tbp]
    \centering
    \includegraphics[width=\columnwidth]{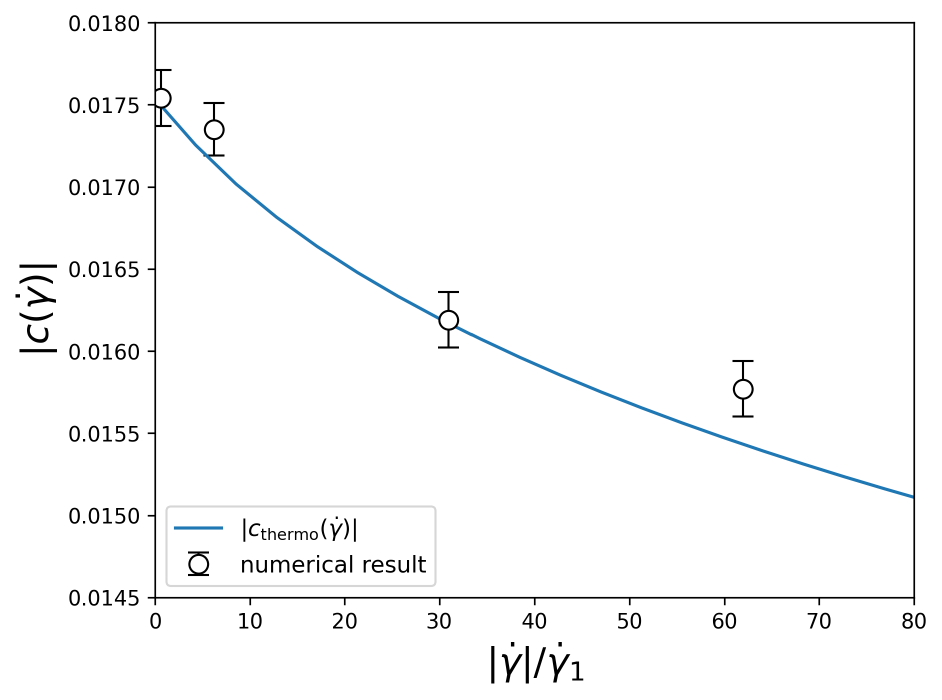}
\caption{
$\dot{\gamma}$ dependence of $c(\dot{\gamma})$ with $\delta f=0$.
The grid spacing is $\Delta x=0.8$, and the other parameters are the same as those in Fig. \ref{fig:image4}.
The circles show the numerical results given by $\langle V(t=6000) \rangle$.
Error bars indicate the standard error of the mean.
The blue solid line shows $c_{\mathrm{thermo}}(\dot{\gamma})$.
}
\label{fig:image5}
\end{figure}

\begin{figure}[tbp]
     \centering
     \includegraphics[width=\columnwidth]{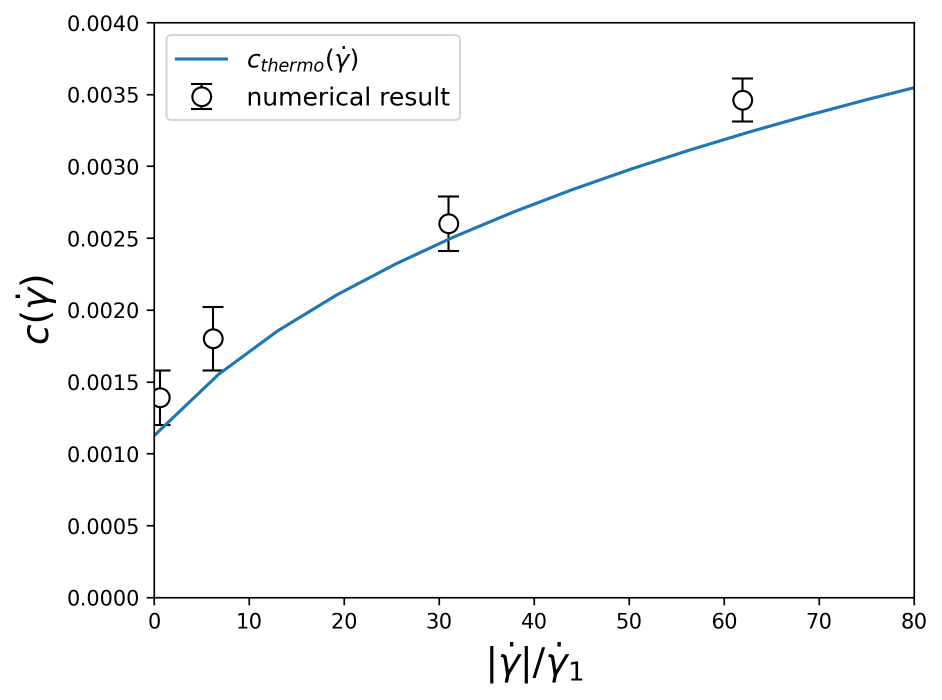}
    \caption{
    $\dot{\gamma}$ dependence of $c(\dot{\gamma})$ with $h=-0.004$, i.e., $\delta f \simeq 0.008$.
    The grid spacing is $\Delta x=1.0$, the system sizes are $L_1=400$ and $L=100$,
    and the other potential parameters are the same as those in Fig. \ref{fig:image4}.
    The circles show the numerical results given by $\langle V(t=12000) \rangle$ from 60 samples.
    Error bars indicate the standard error of the mean.
    The blue solid line shows $c_{\mathrm{thermo}}(\dot{\gamma})$.
    }
  \label{fig:image6}
\end{figure}

Figures \ref{fig:image5} and \ref{fig:image6} show the propagation velocity of the interface, $c(\dot{\gamma})$, as the shear rate is varied for $h=0$, $\delta f=0$, and for $h=-0.004$, $\delta f\simeq0.008$, respectively.
The shear-rate values are normalized by $\dot{\gamma}_1$.
The circles show the numerical results, and the blue solid line shows $c_{\mathrm{thermo}}(\dot{\gamma})$.
The measured velocity broadly follows $c_{\mathrm{thermo}}(\dot{\gamma})$, while small deviations remain at the level of the present numerical uncertainty.
These deviations may include residual contributions such as $g_{\mathrm{neq}}$, but the present simulations do not resolve the shear-rate dependence of $g_{\mathrm{neq}}$. 
We note that the low-shear correction proportional to $|\dot{\gamma}|^{4/3}$ is not resolved because it is smaller than the numerical uncertainty.



\section{Derivation of Eq. (\ref{result})}

We now derive a formula for the propagation velocity of the interface under shear flow.
For deterministic systems, several methods have been developed to derive equations of motion for interfaces \cite{Meron, Cross, Ei, Karma, Hiraizumi}.
The common idea is to extract the interface position as the slow variable while eliminating the faster degrees of freedom.
We assume that $T$ and $h$ are sufficiently small and treat
them as perturbations around $\varphi_0$.
The velocity equation contains the previously derived result for the interface dynamics given in Ref. \cite{Bray}.

\subsection{Setup of the perturbation}

First, we introduce a small parameter and set up the perturbation expansion.
To make the perturbation order explicit, we replace $T$ by $\epsilon^2 T'$ and $h$ by $\epsilon h'$, where $\epsilon$ is a small dimensionless parameter.
We introduce the comoving coordinate $w \equiv x_1-\theta({\bm r}_{\perp}, t)$.
We also define the normalized noise $\tilde{\eta}$ by $\sqrt{2\Gamma T}\tilde{\eta}(w,{\bm r}_\perp,t)\equiv \eta({\bm r},t)$. The normalized noise satisfies
\begin{eqnarray}
\lefteqn{\langle \tilde{\eta}(w,{\bm r}_\perp,t)\tilde{\eta}(w',{\bm r'}_{\perp},t') \rangle}\notag \\
&&= \delta(w-w')\delta({\bm r}_\perp- {\bm r'}_{\perp})\delta(t-t').
\end{eqnarray}
We expand the solution around $\varphi_0$ as
\begin{eqnarray}
\phi({\bm r}, t) &=& \varphi_0(w) \notag \\
&+& \epsilon \rho_1(w,{\bm r}_{\perp}, t)
+ \epsilon^2 \rho_2(w,{\bm r}_{\perp}, t)+O(\epsilon^3).
\end{eqnarray}
Here, $\theta({\bm r}_{\perp},t)$ is the interface position at transverse coordinate ${\bm r}_{\perp}$ and time $t$.
Assuming $\partial_{x_i}\theta$ is small and proportional to $\epsilon$, we introduce a large-scale coordinate ${\bm R}_{\perp}\equiv \epsilon {\bm r}_{\perp}$ and define $\Theta({\bm R}_{\perp}, t) \equiv \theta({\bm r}_{\perp}, t)$.
The time evolution of $\Theta(\bm{R}_{\perp}, t)$ is
\begin{eqnarray}
\partial_t \Theta = \epsilon \Omega_1([\Theta]) + \epsilon^2 \Omega_2([\Theta]) + O(\epsilon^3).
\end{eqnarray}
Here, $[\Theta]$ denotes functional dependence on $\Theta$, $\partial_{X_i} \Theta$, and $\partial_{X_i}^2 \Theta$ for $i=2,\cdots,d$.
The perturbative projection method can be regarded as a stochastic generalization of the approach in Ref. \cite{Kuramoto}.
The quantity of interest is the propagation velocity
$c(\dot{\gamma})\equiv \langle \partial_t\Theta\rangle$.


To simplify the calculation, we next introduce some notation.
We define the linear operator
\begin{eqnarray}
\hat{L}_w\equiv-f_0^{''}(\varphi_0(w))+ \kappa\partial_w^2,
\end{eqnarray}
which describes the linear dynamics of perturbations around $\varphi_0(w)$.
With $u_0(w)\equiv\partial_w \varphi_0(w)$, translational symmetry gives $\hat{L}_w u_0(w)=0$.
Thus, $u_0$ is the Goldstone mode associated with the broken translational symmetry of the interface solution and represents the interface mode.
We also define the projection onto the mode $u_0$ by
\begin{eqnarray}
\hat{P} g \equiv \frac{(u_0, g)}{(u_0, u_0)} u_0,
\label{P}
\end{eqnarray}
for $g(w,{\bm r}_\perp)$, where $(g_1, g_2)$ denotes the inner product
\small
\begin{eqnarray}
(g_1, g_2) \equiv \frac{1}{L^{d-1}} \int_{[0,L]^{d-1}} d{\bm r}_{\perp}
\int_{-\infty}^{\infty} dw\, 
g_1(w,{\bm r}_{\perp})g_2(w,{\bm r}_{\perp})
\end{eqnarray}
\normalsize
for functions $g_1(w, {\bm r}_{\perp})$ and $g_2(w, {\bm r}_{\perp})$ defined on
$[-\infty, \infty] \times [0, L]^{d-1}$.
Furthermore, $\hat{Q} \equiv 1 - \hat{P}$.


Finally, we use the solvability condition to derive the result of the perturbation expansion.
At first order in the perturbation expansion, we obtain
\begin{eqnarray}
(\partial_{t} - \Gamma(\hat{L}_w + \kappa \Delta_{{\bm r}_{\perp}}))\rho_1=B_1,
\label{rho_1_B}
\end{eqnarray}
\small
\begin{eqnarray}
B_1&\equiv&\Omega_1u_0+\Gamma h'  \notag \\
&&- \dot{\gamma}(w+\Theta)(-\partial_{X_2}\Theta u_0 + \partial_{x_2}\rho_1)+\sqrt{2\Gamma T'} \tilde{\eta}.
\end{eqnarray}
\normalsize
Here, we set $\Delta_{{\bm r}_{\perp}}\equiv\partial_{x_2}^2+\cdots+\partial_{x_d}^2$.
Because 
\begin{eqnarray}
(\partial_t - \Gamma(\hat{L}_w + \kappa \Delta_{{\bm r}_\perp}) )u_0 = 0,
\end{eqnarray}
the equation for $\rho_1$ has a solution only if $(u_0, B_1) = 0$.
This solvability condition determines both $\Omega_1$ and the equation of motion for $\rho_1$.
The second-order terms are obtained in the same way.

\subsection{Result of the perturbation expansion}

The first-order solvability condition gives
\small
\begin{eqnarray}
\Omega_1([\Theta])&=&-\frac{\Gamma h'\Delta\phi}{(u_0,u_0)}
-\dot{\gamma}(\Theta+a)\partial_{X_2} \Theta \notag \\
&&- \frac{(u_0,\sqrt{2\Gamma T'}\tilde{\eta})}{(u_0,u_0)},
\end{eqnarray}
\normalsize
\begin{eqnarray}
&&\left(\partial_{t}+\dot{\gamma}(w+\Theta)\partial_{x_2}
- \Gamma(\hat{L}_{w} + \kappa \Delta_{{\bm r}_{\perp}})\right)\rho_1 \notag \\
&=& \dot{\gamma} (w-a)\partial_{X_2} \Theta u_0(w) +\Gamma \hat{Q}h' + \sqrt{2\Gamma T'} \hat{Q}\tilde{\eta}.
\label{rho1}
\end{eqnarray}
Here, $a \equiv (u_0, wu_0)/(u_0, u_0)$.
Next, the second-order solvability condition leads to
\small
\begin{eqnarray}
\Omega_2([\Theta])&=&
\Gamma \kappa \Delta_{{\bm R}_{\perp}} \Theta
+ \frac{\Gamma (u_0, \frac{1}{2} f_0^{'''}(\varphi_0)\rho_1^2)}{(u_0,u_0)} \notag \\
&&- \Omega_1([\Theta])\frac{(u_0, \partial_{w} \rho_1)}{(u_0,u_0)}
- \dot{\gamma} \partial_{{X}_2} \Theta
\frac{(u_0, w \partial_{w}\rho_1)}{(u_0,u_0)} \notag \\
&&- \dot{\gamma} \Theta \partial_{{X}_2} \Theta
\frac{(u_0, \partial_{w}\rho_1)}{(u_0,u_0)}.
\end{eqnarray}
\normalsize
Since the advection and diffusion terms do not contribute to the driving force, we omit them in the expression for the velocity.
Setting $\epsilon\to1,T'\to T,h'\to h$,
\begin{eqnarray}
c(\dot{\gamma}) = -\frac{\Gamma h \Delta \phi}{(u_0,u_0)} &+& \frac{\Gamma (u_0, \frac{1}{2} f_0^{'''}(\varphi_0)\langle \rho_1^2 \rangle)}{(u_0,u_0)} \notag \\
+ \frac{\Gamma h \Delta \phi \langle (u_0,\partial_w \rho_1) \rangle}{(u_0,u_0)^2} &+& O(T^{\frac{3}{2}},Th,T^{\frac{1}{2}}h^2,h^3),
\label{c_form}
\end{eqnarray}
which extends the result of Ref. \cite{Iwata}.
The derivation of Eq. (\ref{c_form}) is presented in Appendix \ref{app:velocity_formula}.
We neglect the third term, as it is $O(hT^{\frac{1}{2}})$.

\subsection{Derivation of the driving force}

We now derive Eq. (\ref{result}) from Eq. (\ref{c_form}).
The second term in Eq. (\ref{c_form}) can be rewritten as
\begin{multline}
\frac{1}{L^{d-1}} \int_{[0,L]^{d-1}} d{\bm r}_{\perp}\int_{-\infty}^{\infty} dw\,
\frac{1}{2} u_0 f_0^{'''}(\varphi_0) \langle \rho_1^2 \rangle \\
= \lim_{\Lambda \to \infty}[ \Psi[w;\rho_1(w)] ]|_{w=-\Lambda}^{w=\Lambda}
+ g_{\mathrm{neq}} \\
- h^2 \Big[ \frac{1}{f^{''}(\varphi_2)} - \frac{1}{f^{''}(\varphi_1)} \Big] + \frac{h \Delta \phi}{(u_0,u_0)}
\langle(\partial_w \rho_1,u_0) \rangle.
\label{driving_force}
\end{multline}
Here,
\begin{eqnarray}
\Psi[w;\rho_1(w)]&\equiv&\frac{1}{2}\big[f_0^{''}(\varphi_0(w)) \langle \rho_1(w)^2 \rangle \notag \\
&&-\kappa \langle (\partial_w \rho_1)^2 \rangle
+ \kappa \langle (\nabla_{{\bm r}_{\perp}}\rho_1)^2 \rangle\big],
\label{psi}
\end{eqnarray}
\begin{eqnarray}
g_{\mathrm{neq}} &\equiv& \frac{1}{\Gamma L^{d-1}} \int_{[0,L]^{d-1}} d{\bm r}_{\perp} \int_{-\infty}^{\infty} dw \notag \\
&&\left\langle \partial_w \rho_1
\left(\partial_{t}+\dot{\gamma}(w+\Theta({\bm R}_{\perp},t))\partial_{x_2}\right)\rho_1\right\rangle.
\end{eqnarray}
The derivation of Eq. (\ref{driving_force}) is presented in Appendix \ref{app:velocity_formula}.
In the bulk region $w=\mu_i \Lambda$, we decompose $\rho_1$ as
\begin{eqnarray}
\rho_1(w=\mu_i \Lambda)=\frac{h}{f^{''}(\varphi_i)} + \delta \rho_1(\mu_i \Lambda),
\end{eqnarray}
and define
\begin{eqnarray}
-Ts_i(\dot{\gamma})= \lim_{\Lambda \to \infty}\Psi[\mu_i \Lambda; \delta \rho_1(\mu_i \Lambda)].
\label{psi_delta}
\end{eqnarray}
Here, $\mu_i=2i-3,i=1,2$.
In Appendix \ref{app:velocity_formula}, the expectation value in Eq. (\ref{psi_delta}) is approximated by the corresponding expectation value for a linearized potential.
Neglecting terms of order $O(hT^{\frac{1}{2}},h^2)$ then gives Eq. (\ref{result}).

\section{Conclusion}

We derive a formula for the propagation velocity of a planar interface driven by an entropic force under shear flow.
The driving force consists of a shear-modified entropic contribution and a contribution arising from the breaking of time-reversal symmetry.
In the low-shear-rate regime and for small free-energy bias, the leading shear-induced correction to the velocity of the interface near $x_1=0$ is proportional to the $4/3$ power of the shear rate.
In the high-shear-rate regime, numerical simulations in $d=2$ show that the measured velocities are largely captured by the shear-modified bulk entropic contribution appearing in the velocity formula.
These results suggest that shear flow can alter the propagation velocity through the entropic force.
A natural next step is therefore to investigate the present dynamics at a more microscopic level and to examine whether $g_{\mathrm{neq}}$ emerges and how it depends on $\dot{\gamma}$.

Experimental verification remains an important challenge.
Liquid-crystal systems under shear provide a natural setting for observing the propagation phenomenon studied here, 
because shear-induced changes in orientational and phase behavior have been observed \cite{Pujolle-Robic,Mendil-Jakani,D_Parisi},
although liquid-crystal systems are not direct realizations of the present scalar Model A dynamics.
In the present theory, the characteristic shear rates defining the low-shear regime satisfy $\dot{\gamma}_1\sim\dot{\gamma}_2\sim\Gamma$.
If $\Gamma$ is estimated from the orientational relaxation scale associated with rotational viscosity in liquid-crystal systems \cite{Stallinga,Kim}, this condition suggests that $\dot{\gamma}_1$ can be on the order of $10^{-2}\,{\rm s}^{-1}$,
indicating that the relevant shear-rate scale may be experimentally accessible.
Parallel-plate shear cells \cite{Wu_2007} and rotating-cylinder geometries \cite{Kang} are therefore promising platforms for such tests.
At the same time, quantitatively measuring the propagation velocity of the interface requires spatial resolution comparable to the interface width, which is challenging in molecular fluids, polymer blends, and ferroelectric or ferroelastic materials,
where interfacial profiles or domain-wall widths can be on the nanometer scale \cite{Townsend,Binder_2001,Catalan_2012}.
Experimental platforms with nanoscale spatial resolution, such as magnetic-domain-wall systems \cite{Yamaguchi,Hayashi,Ravelosona},
may therefore offer a complementary route to measure the propagation velocity.

Several theoretical extensions are also important.
First, the present model uses a non-conserved order parameter, whereas many phase-separating systems involve conserved fields such as density or concentration \cite{Cahn_Hilliard,Langer,Onuki_2002}.
Extending our theory to interface dynamics of conserved systems \cite{Kawasaki_1982,Ohta_1984,Bray,Sarfati} would clarify the properties of domain growth and interface fluctuations in non-equilibrium systems.
Second, the present analysis treats a flat interface, while real interfaces generally exhibit surface topography and fluctuation-induced roughening \cite{KPZ,Halpin-Healy,Ramsay}.
Deriving an interface equation that includes both interfacial topography and coupling to fluid flow would connect the present mechanism to broader theories of fluctuating interfaces \cite{KPZ,Halpin-Healy,Ramsay}.
Third, it remains to understand how shear-modified entropic driving affects nucleation barriers and growth laws \cite{Langer_1969,Langer,Onuki_2002}.
Finally, the relation between the velocity formula and entropy production in non-equilibrium steady states should be clarified \cite{Komatsu,Bertini,Maes,Seifert_2012,Seifert_2019,Carosi}.
Establishing this relation may yield quantitative constraints or bounds on fluctuation-driven interfacial motion in non-equilibrium steady states.

\begin{acknowledgments}
This study was supported by JSPS KAKENHI Grant Numbers JP25K00923 and JP26H00383, JST SPRING Grant Number JPMJSP2110, and Grant-in-Aid for JSPS Fellows Grant No. 26KJ1403.
\end{acknowledgments}


\appendix

\section{Supplementary information on the derivation}
\label{app:derivation}

This appendix provides the technical details behind the velocity formula used in the main text.
It is organized as follows.
In Sec. \ref{app:velocity_formula}, we derive the velocity formula by projecting the perturbative interface dynamics onto the translational zero mode and separating the fluctuation contribution into a boundary term and the remaining non-equilibrium term $g_{\mathrm{neq}}$.
In Sec. \ref{linear_system}, we derive the homogeneous-bulk fluctuation correlations used to analyze $s_i(\dot{\gamma})$ and $g_{\mathrm{neq}}$ in the main text.

\subsection{Derivation of the main result}
\label{app:velocity_formula}

This subsection derives the velocity formula from the perturbative interface dynamics in Sec. IV.
We first project the dynamics onto the translational zero mode to obtain Eq. (\ref{c_form}).
We then decompose the fluctuation term in Eq. (\ref{c_form}) into the boundary contribution, which becomes the shear-modified entropic term, and the remaining non-equilibrium contribution $g_{\mathrm{neq}}$.
All averages are taken over noise realizations in the perturbative dynamics.

\subsubsection{Derivation of Eq. (\ref{c_form})}
\label{derivation_c_form}

We derive Eq. (\ref{c_form}) in the main text by projecting the perturbative dynamics onto the translational zero mode of the interface.
Keeping terms up to second order in $\epsilon$ gives
\begin{eqnarray}
c(\dot{\gamma}) = \langle \Omega_1([\Theta]) \rangle + \langle \Omega_2([\Theta])\rangle.
\end{eqnarray}
The first-order contribution is $\langle \Omega_1([\Theta]) \rangle$.
The advection term does not contribute after averaging, and the noise average vanishes because $\langle \eta \rangle=0$.
Thus,
\begin{eqnarray}
\langle \Omega_1([\Theta]) \rangle=-\frac{\Gamma h \Delta \phi}{(u_0,u_0)}.
\end{eqnarray}
\begin{widetext}
The second-order contribution is evaluated in the same way.
Combining it with the first-order contribution yields
\begin{eqnarray}
c(\dot{\gamma}) &\equiv& -\frac{\Gamma h \Delta \phi}{(u_0,u_0)} + \frac{\Gamma ( u_0, \frac{1}{2} f_0^{'''}(\varphi_0)\langle \rho_1^2 \rangle )}{(u_0,u_0)} - \frac{\langle \Omega_1([\Theta])(u_0, \partial_{w} \rho_1) \rangle}{(u_0,u_0)}.
\label{c_form_1}
\end{eqnarray}
The third term in Eq. (\ref{c_form_1}) can be written as
\begin{eqnarray}
\frac{\langle \Omega_1([\Theta])(u_0, \partial_{w} \rho_1) \rangle}
{(u_0,u_0)}
&=& -\frac{\Gamma h \Delta \phi}{(u_0,u_0)^2}
\langle(u_0, \partial_{w} \rho_1) \rangle
- \frac{\sqrt{2\Gamma T} }{(u_0,u_0)}
\langle (u_0,\tilde{\eta})(u_0, \partial_{w} \rho_1) \rangle.
\label{c_form_2}
\end{eqnarray}
We next show that the second term in Eq. (\ref{c_form_2}) vanishes.
In the noise-correlation calculation below,
${\bm r}_{\perp 1}$ and ${\bm r}_{\perp 2}$ are integrated over
$[0,L]^{d-1}$, while $w_1$ and $w_2$ are integrated over
$(-\infty,\infty)$.
\begin{eqnarray}
 &\sqrt{2\Gamma T}& (L^{d-1})^2 \langle (u_0,\tilde{\eta})(u_0, \partial_{w} \rho_1) \rangle \notag \\
&=&  \sqrt{2\Gamma T}  
\int d {\bm r}_{\perp 1} d {\bm r}_{\perp 2} dw_1 dw_2 \,
u_0(w_1) u_0(w_2)
\langle \tilde{\eta}(w_1, {\bm r}_{\perp 1})\partial_{w_2} \rho_1(w_2, {\bm r}_{\perp 2}) \rangle \notag \\
&=& 2\Gamma T \int d {\bm r}_{\perp 1} d {\bm r}_{\perp 2} dw_1 dw_2 u_0(w_1) u_0(w_2)
\partial_{w_2} \Big[\delta(w_1-w_2)\delta({\bm r}_{\perp 1}-{\bm r}_{\perp 2})
-\frac{ u_0(w_1) u_0(w_2)}{(u_0,u_0)L^{d-1}} \Big] \notag \\
&=&0.
\label{c_form_3}
\end{eqnarray}
In this calculation, the noise correlation is evaluated in the Stratonovich convention.
Combining Eqs. (\ref{c_form_1}), (\ref{c_form_2}), and (\ref{c_form_3}) gives Eq. (\ref{c_form}) in the main text.

\subsubsection{Derivation of Eq. (\ref{driving_force})}
\label{derivation_result}
We next derive Eq. (\ref{driving_force}) in the main text.
Using the relation
\begin{eqnarray}
u_0(w)f_0^{'''}(\varphi_0(w))
=\frac{ d f_0^{''}(\varphi_0(w)) }{dw},
\end{eqnarray}
we obtain
\begin{eqnarray}
\int_{-\infty}^{\infty} dw \frac{1}{2}u_0 f_0^{'''}(\varphi_0(w)) \rho_1^2
= \left[\frac{1}{2} f_0^{''}(\varphi_0(w))\rho_1^2 - \frac{\kappa}{2}(\partial_w \rho_1)^2 \right]_{-\infty}^{\infty} + \int_{-\infty}^{\infty} dw \partial_w \rho_1 \hat{L}_w \rho_1.
\label{driving_force_1}
\end{eqnarray}
Next, we integrate both sides of Eq. (\ref{driving_force_1}) with respect to ${\bm r}_{\perp}$.
We then evaluate the second term on the right-hand side of Eq. (\ref{driving_force_1}).
By multiplying Eq. (\ref{rho1}) in the main text by $\partial_w \rho_1$, this term can be evaluated as follows:
\begin{eqnarray}
\frac{\Gamma}{L^{d-1}} \int d {\bm r}_{\perp }
\int dw  \partial_w \rho_1 \hat{L}_w \rho_1 
&=& \Gamma \left[\frac{\kappa}{2}
 (\nabla_{{\bm r}_{\perp} }\rho_1)^2 
\right]_{w=-\infty}^{w=\infty} \notag \\
&&+ \frac{1}{L^{d-1}} \int d {\bm r}_{\perp } \int dw
 \partial_w \rho_1
(\partial_{t}+\dot{\gamma}(w+\Theta({\bm R}_{\perp}))
\partial_{x_2})\rho_1 \notag \\
&&-(\partial_w \rho_1,\Gamma \hat{Q}h) - \partial_{X_2} \Theta (\partial_w \rho_1, \dot{\gamma}(w-a)u_0)- (\partial_w \rho_1, \sqrt{2\Gamma T} \hat{Q}\tilde{\eta}).
\label{driving_force_2}
\end{eqnarray}
The first term in Eq. (\ref{driving_force_2}) is obtained by partial integration with respect to ${\bm r}_{\perp }$.
We omit the fourth term in Eq. (\ref{driving_force_2}) because this term represents advection and therefore does not contribute to interfacial propagation.
Finally, we average Eq. (\ref{driving_force_2})  over the noise.
The third term in Eq. (\ref{driving_force_2}) is calculated as
\begin{eqnarray}
\langle (\partial_w \rho_1,\Gamma \hat{Q}h) \rangle 
&=& \frac{\Gamma }{L^{d-1}}\int d {\bm r}_{\perp } \int dw \Big[ \partial_w \langle \rho_1 \rangle h  -\frac{(u_0,h)}{(u_0,u_0)} \partial_w \rho_1 u_0 \Big] \notag \\
&=& \Gamma h \Big[\frac{h}{f^{''}(\phi_2)}- \frac{h}{f^{''}(\phi_1)}\Big] - \frac{\Gamma h \Delta \phi}{(u_0,u_0)}  \langle (\partial_w \rho_1,u_0) \rangle .
\label{Qh}
\end{eqnarray}
\end{widetext}
Here, we used Eq. (\ref{rho1}), which implies that $\rho_1$ approaches $h/f^{''}(\varphi_1)$ and $h/f^{''}(\varphi_2)$ in the two bulk regions.

\subsection{Bulk fluctuation correlations under shear flow}
\label{linear_system}

In this subsection, we derive two identities for homogeneous bulk fluctuations that are used in the main text.
First, we obtain the shear-modified equal-time correlation function, which determines $s_i(\dot{\gamma})$ for $i=1,2$.
Second, we show that the local correlation involving the convective time derivative vanishes in the homogeneous steady state.
This result explains why the integrand of $g_{\mathrm{neq}}$ is localized near the interface.
A linearized bulk mode is sufficient for these purposes because the entropic contribution is determined by the Gaussian fluctuation spectrum around each homogeneous phase.
Let the spatial coordinates be ${\bm r}=(x_1,\cdots,x_d)$.
We consider a fluctuation field $\psi$ around a uniform bulk state, governed by
\begin{eqnarray}
(\partial_{t} + \dot{\gamma}x_1\partial_{x_2}
+ \Gamma[r - \kappa\Delta ])\psi
=  \Gamma h + \sqrt{2 \Gamma T}\zeta, \\
\langle \zeta({\boldsymbol r},t) \zeta({\boldsymbol r'},t') \rangle = \delta({\boldsymbol r} -{\boldsymbol r'})\delta(t-t').
\end{eqnarray}
The constant term $\Gamma h$ only shifts the spatially uniform mean value of the field.
Writing
\begin{eqnarray}
\psi({\bm r},t)=\bar{\psi}+\delta\psi({\bm r},t), \qquad \bar{\psi}=\frac{h}{r},
\end{eqnarray}
the centered fluctuation $\delta\psi$ satisfies
\begin{eqnarray}
(\partial_{t} + \dot{\gamma}x_1\partial_{x_2}
+ \Gamma[r - \kappa\Delta ])\delta\psi
= \sqrt{2 \Gamma T}\zeta.
\label{bulk_centered}
\end{eqnarray}
Since both $s_i(\dot{\gamma})$ and Eq. (\ref{rho_1_time}) depend only on fluctuation correlations, the following calculation is carried out for $\delta\psi$.
To avoid cumbersome notation, we denote $\delta\psi$ simply by $\psi$ below.
We define the Fourier transform by
\begin{eqnarray}
\hat{\psi}({\bm k},t)= \int d {\bm r} \delta \psi({\bm r},t) e^{i {\bm k}{\bm r} }.
\end{eqnarray}
The equal-time correlation function is defined by
\begin{eqnarray}
C({\bm k},t)\delta({\bm k}+{\bm k}') \equiv \langle \hat{\psi}({\bm k},t) \hat{\psi}({\bm k}',t) \rangle.
\end{eqnarray}
The steady-state correlation function is then approximated by
\begin{eqnarray}
 C({\bm k}) \simeq \frac{(2\pi)^d T}{
 r+\kappa {\bm k}^2
 + a_0(\Gamma^{-1}\sqrt{\kappa}|\dot{\gamma}||k_2|)^{2/3} }.
 \label{form_C}
\end{eqnarray}
The steady-state correlation function also implies that
\begin{eqnarray}
\left\langle (\partial_{t} + \dot{\gamma}x_1\partial_{x_2})
\psi({\bm r}) \partial_{x_1} \psi({\bm r}) \right\rangle = 0.
\label{rho_1_time}
\end{eqnarray}
Equation (\ref{rho_1_time}) is used below to discuss the non-equilibrium contribution.

\subsubsection{Derivation of Eq. (\ref{form_C})}
We derive Eq. (\ref{form_C}) by following the characteristic curves in wave-number space.
The characteristic-curve representation is useful because the shear term couples different wave-number components.
Taking the Fourier transform of the linearized dynamics gives
\begin{eqnarray}
\partial_t \hat{\psi} - \dot{\gamma} k_2 \partial_{k_1} \hat{\psi}
&=& - \Gamma (r + \kappa {\bm k}^2)\hat{\psi}
+ \sqrt{2\Gamma T}\hat{\zeta}.
\end{eqnarray}
The noise correlation is
\begin{eqnarray}
\langle \hat{\zeta}({\bm k},t)\hat{\zeta}({\bm k}',t') \rangle
=(2\pi)^d\delta({\bm k}+{\bm k}')\delta(t-t').
\end{eqnarray}
To solve this equation, we introduce the characteristic variables
\begin{eqnarray}
t(s)&=&t+s, \\
k_1(s)&=& k_1 - \dot{\gamma} k_2 s.
\end{eqnarray}
For $\psi(k_1,{\bm k}_{\perp},s)\equiv\hat{\psi}(k_1(s),{\bm k}_{\perp},t(s))$, the evolution equation becomes
\begin{eqnarray}
\partial_s \hat{\psi} (k_1,{\bm k}_{\perp},s)
&=&-\Gamma[r+\kappa(k_1(s)^2+{\bm k}_{\perp}^2)]
\hat{\psi} (k_1,{\bm k}_{\perp},s) \notag \\
&&+\sqrt{2\Gamma T}\hat{\zeta}(k_1,{\bm k}_{\perp},s).
\end{eqnarray}
Solving this equation and evaluating the equal-time correlation function gives
\begin{widetext}
\begin{eqnarray}
\langle \hat{\psi} ({\bm k},t) \hat{\psi} ({\bm k}',t) \rangle 
&=&\langle \hat{\psi} ({\bm k},s=0) \hat{\psi} ({\bm k}',s=0) \rangle \notag \\
&=& 2(2\pi)^d \Gamma T \int_{-\infty}^{0} ds
\,{\rm exp}\left( 2\Gamma\left[ (r+\kappa {\bm k}^2)s - \dot{\gamma} \kappa k_1 k_2 s^2
+ \frac{1}{3} \dot{\gamma}^2 \kappa k_2^2 s^3 \right]\right)\delta({\bm k}+{\bm k}') \notag \\
&=&(2\pi)^d \Gamma T \int_{0}^{\infty} ds
\,{\rm exp}\bigl(-\Gamma[(r+\kappa {\bm k}^2)s
+ \frac{1}{2}\dot{\gamma} \kappa k_1k_2s^2
+ \frac{1}{12}\dot{\gamma}^2\kappa k_2^2 s^3]\bigr)\delta({\bm k}+{\bm k}').
\end{eqnarray}
Evaluating the integral as in Ref. \cite{Nakano} yields
\begin{eqnarray}
 C({\bm k}) \simeq \frac{(2\pi)^d T}{r+\kappa {\bm k}^2 + a_0(\Gamma^{-1}\sqrt{\kappa}|\dot{\gamma}||k_2|)^{2/3} }.
\end{eqnarray}

\subsubsection{Derivation of Eq. (\ref{rho_1_time})}
We next derive Eq. (\ref{rho_1_time}).
Substituting $\psi(k_1,{\bm k}_{\perp},s)\equiv\hat{\psi}(k_1(s),{\bm k}_{\perp},t(s))$ into the left-hand side of Eq. (\ref{rho_1_time}), we obtain
\begin{eqnarray}
\langle (\partial_t &+& \dot{\gamma}x_1 \partial_{x_2}) \psi({\bm r})  \partial_{x_1}\psi({\bm r})  \rangle 
= \int \frac{d {\bm k}}{(2\pi)^d}
\langle (\partial_t  - \dot{\gamma} k_2 \partial_{k_1} )
\hat{\psi}({\bm k}) (-i k_1) \hat{\psi}(-{\bm k}) \rangle \notag \\
&=&\int \frac{d {\bm k}}{(2\pi)^d}
\langle  \partial_s \hat{\psi}({\bm k},s=0)
(-i k_1) \hat{\psi}(-{\bm k},s=0) \rangle \notag \\
&=&- \int \frac{d {\bm k}}{(2\pi)^d}
\Gamma[r+\kappa{\bm k}^2](-ik_1 )
\langle \hat{\psi} ({\bm k},0)\hat{\psi} (-{\bm k},0) \rangle
+ \int \frac{d {\bm k}}{(2\pi)^d}
\sqrt{2\Gamma T} (-ik_1 )
\langle \hat{\zeta}({\bm k},0) \hat{\psi} (-{\bm k},0) \rangle.
\end{eqnarray}
The first contribution vanishes because
\begin{eqnarray}
\int \frac{d {\bm k}}{(2\pi)^d}
&& \Gamma[r+\kappa{\bm k}^2] (-ik_1)
\langle \hat{\psi} ({\bm k},0)\hat{\psi} (-{\bm k},0) \rangle
\notag \\
&=&\int \frac{d {\bm k}}{(2\pi)^d}
\Gamma[r+\kappa{\bm k}^2] (-i k_1) (2\pi)^d \Gamma T \int_{0}^{\infty} ds
{\rm exp}\left(-\Gamma\left[ (r+\kappa {\bm k}^2)s
+ \frac{1}{2} \dot{\gamma} \kappa k_1 k_2 s^2
+ \frac{1}{12} \dot{\gamma}^2 \kappa k_2^2 s^3 \right]\right)
\notag \\
&=& (2\pi)^d \Gamma T \int_{0}^{\infty} ds \int \frac{d k_1}{2\pi}
\int_0^{\infty} \frac{d k_2}{2\pi} \cdots \frac{d k_d}{2\pi}
\Gamma[r+\kappa {\bm k}^2] (-i k_1) \notag \\
&& \times {\rm exp}\left(-\Gamma\left[ (r+\kappa {\bm k}^2)s
+ \frac{1}{12} \dot{\gamma}^2 \kappa k_2^2 s^3 \right]\right)
2 {\rm cosh}\left(\frac{1}{2} \dot{\gamma}\kappa k_1 k_2 s^2\right) \notag \\
&=&0.
\end{eqnarray}
The noise contribution vanishes by the same symmetry argument.
The subtracted uniform mean $\bar{\psi}=h/r$ does not contribute either, because $(\partial_t+\dot{\gamma}x_1\partial_{x_2})\bar{\psi}=0$ and $\partial_{x_1}\bar{\psi}=0$.
Thus Eq. (\ref{rho_1_time}) is proved.

\end{widetext}

\section{Numerical simulations}
\label{app:numerics}
We provide additional details on the numerical computations.

\subsection{Numerical setup}

We define the discrete model on a square grid with spatial mesh size $\Delta x$, chosen to be smaller than both $\xi_1$ and $\xi_2$.
The model is obtained by discretizing Eq. (\ref{dynamics}), with $\partial_x^2 + \partial_y^2$ replaced by the finite-difference Laplacian.
All numerical calculations are performed for $d=2$.
To avoid numerical divergence associated with the unbounded shear velocity as $x \to \pm \infty$, we introduce the coordinate transformation
\begin{eqnarray}
x'&=&x, \\
y'&=&y-\dot{\gamma}tx, \\
t'&=&t.
\end{eqnarray}
For the transformed field $\varphi({\bm r}',t')=\phi({\bm r},t)$, the dynamics becomes
\begin{eqnarray}
\partial_t \varphi({\bm r}',t')
&=& -\Gamma[f'(\varphi)
- \kappa((\partial_x- \dot{\gamma}t\partial_y)^2+\partial_y^2)\varphi]
\notag \\
&&+\sqrt{2\Gamma T} \eta({\bm r}',t').
\label{varphi_dynamics}
\end{eqnarray}
The stochastic time evolution is computed with the Heun method.
During the integration, the coordinate transformation is periodically reset so that the accumulated shear displacement remains bounded.
The procedure is:

1. Transform $\phi({\bm r},0)$ into $\varphi({\bm r}',t')$ at $t=0$.

2. Solve Eq. (\ref{varphi_dynamics}) up to $t'=1/\dot{\gamma}$.

3. Transform $\varphi({\bm r}',t')$ into $\phi({\bm r},t)$ at $t'=1/\dot{\gamma}$.

4. Reset $t'$ from $1/\dot{\gamma}$ to $0$ and return to step 1.

\noindent
This reset procedure follows Refs. \cite{Toh,Onuki_1997}.


\end{document}